\documentclass[aps,prd,twocolumn,groupedaddress,nofootinbib]{revtex4}
\usepackage{graphics}

\usepackage{latexsym}
\usepackage{amsmath,amssymb}

\begin{document}
\title{Gravitational inflaton decay and the hierarchy problem}
\author{Yuki Watanabe}
\email{yuki@astro.as.utexas.edu}
\affiliation{Department of Physics, University of Texas, Austin, Texas
78712, USA} 
\author{Eiichiro Komatsu}
\affiliation{Department of Astronomy, University of Texas, Austin, Texas
78712, USA} 
\date{\today}

\begin{abstract}
We study implications of the large-$N$ species solution to the
 hierarchy problem, proposed by G. Dvali, for reheating of the universe
 after inflation. Dvali's proposal contains additional $N\sim
 10^{32}\;Z_2$-conserved quantum fields beyond the Standard Model
 particles with mass $\sim 1$ TeV, which weaken gravity by a factor of
 $1/N$, and thus explain the hierarchy between the Plank scale and
 the electroweak scale.
We show that, in this scenario, the decay rates of inflaton fields
 through gravitational decay channels are
 enhanced by a factor of $N$, and thus they decay into $N$ species of
 the quantum fields very efficiently, in the limit that 
 quantum gravity effects are unimportant for the gravitational decay rate.
In order not to violate energy conservation or over-reheat the
 universe,  inflaton mass, vacuum expectation value of inflaton, or
 non-minimal gravitational coupling should be tightly fine-tuned. Our
 conclusion holds even when the gravitational decay is prohibited
 by some symmetry of the theory; 
 the universe may still be over-reheated via annihilation of
 inflatons, if the number  density of inflaton quanta is greater than
 the critical value. 
\end{abstract}

\maketitle

\section{introduction}
Gravity is $10^{33}$ times weaker than the weak force. 
Both forces
involve seemingly fundamental constants: Fermi's constant, $G_{\rm
F}=1.17\times 10^{-5}\;{\rm GeV}^{-2}=(293\;{\rm GeV})^{-2}\equiv M_{\rm
w}^{-2}$, for the weak force, and Newton's constant, $G=0.671\times
10^{-38}\;{\rm GeV}^{-2}=(1.22\times 10^{19}\;{\rm GeV})^{-2}\equiv
(\sqrt{8\pi} M_{\rm Pl})^{-2}$ for gravity. 
This inexplicably large
separation between the Planck scale, $M_{\rm Pl}$, and the weak scale,
$M_{\rm w}$, is the so-called gauge hierarchy problem \cite{susskind:82}. 

Why is it a problem? The radiative corrections to the
Higgs boson mass are quadratically divergent and sensitive to the ultraviolet
cutoff scale of the particle physics theory, $M_{\rm UV}$, and thus
drive the bare Higgs mass to a very large value, unless $M_{\rm UV}$ is closer to
$M_{\rm w}$. However, if $M_{\rm UV}$ is not much higher than $M_{\rm w}$,
one may wonder why $M_{\rm Pl}$ is much higher than $M_{\rm UV}$, i.e.,
gravity is so weak.\footnote{In supersymmetric theory,
radiative corrections are only logarithmically divergent. If the
supersymmetry breaking scale is close to $M_{\rm w}$, e.g., $\sim 1$
TeV, it solves the gauge hierarchy problem; however, one still needs to
understand the reason why $M_{\rm Pl}$ and the symmetry breaking scale
are so different, i.e., the $\mu$-problem \cite{QFT3}.}

Dvali recently proposed a simple but radical solution to the hierarchy
problem \cite{dvali:07}. He does not use technicolor or
supersymmetry, but uses the black hole physics to show that any
consistent theory that includes $N Z_2$-conserved species of the quantum
fields with mass $\Lambda$ must have a value of the Planck mass, which is
bounded from below: 
\begin{eqnarray}
\label{dvali_bound}
M_{\rm Pl}^2 \gtrsim N{\Lambda}^2,
\end{eqnarray}
 in a large-$N$ limit.\footnote{The reference \cite{dvali:07} uses the
 Planck mass, $m_{\rm Pl}\sim 10^{19}$ GeV, while we shall use the
 reduced Planck mass, $M_{\rm Pl}\sim 10^{18}$ GeV.} 

Therefore, according to Dvali's solution (see also \cite{veneziano:02}),
gravity is weak because there are
 $N$ species of the quantum fields beyond the Standard Model with mass
 $\Lambda = O({\rm TeV})$, as well as a discrete $Z_2^N$-symmetry, with $N\sim
 10^{32}$. 
An example of this scenario is the celebrated large-extra-dimensions
solution to the hierarchy problem \cite{ADD} (see also \cite{cremades/etal:02} in the context of String Theory), in which $N\sim 10^{32}$
Kaluza-Klein particles of mass $\sim 1$~TeV would appear. 

Can we construct a cosmological model that is consistent with Dvali's
solution to the hierarchy problem? In particular, can we still construct a
successful inflationary scenario, in the presence of such an extremely
large number of extra species at the TeV scale? 

In this paper, we show that Dvali's proposal is consistent with
inflation only when at least one of the followings is tightly fine-tuned:
the inflaton mass, $m_{\sigma}$, vacuum expectation value,
$\langle\phi\rangle\equiv v$, or non-minimal coupling parameter, $\xi$.
While we consider only single field inflation models, our argument can be
extended to a multi-field case in a straightforward manner.
Here, $\phi$ denotes the inflaton field and its mass is
given by the shape of the inflaton potential at the {\it minimum};
$\partial^2V(\phi)/\partial\phi^2|_{\phi=v}\equiv m_{\sigma}^2$. 

Our argument is based exclusively on reheating of the universe after
inflation. Even if
we do not know details of interactions between the inflaton and matter
sector, we do know that there must be decay channels via gravitational
interactions, which give the lower bound of the inflaton decay rate.  

The existence of a large number of quantum fields with mass,
$\Lambda\sim 1$ TeV, will enhance the decay rate of the inflaton field
by $N\sim 10^{32}$. Such a drastic enhancement of the decay rate ought to affect reheating after inflation. 

We shall take a particular point of view when we study implications of
Dvali's proposal. In general, one may consider two cut-off scales: one
for the particle physics, $M_{\rm UV}$, and the other for gravity,
$M_{\rm grav}$, which
 may or may not be the same. In our analysis, we shall assume
 $M_{\rm UV}\sim \Lambda$ and $M_{\rm grav}\gg \Lambda$. 
This assumption allows us to analyze the gravitational inflaton decay in
the semi-classical limit, without worrying about quantum gravitational effects.

The paper is organized as follows. In Sec.~\ref{sec:grav_decay} we
discuss generic properties of the enhanced decay of inflaton. We
especially consider $f(\phi)R$ gravity as an example.
In Sec.~\ref{sec:nonminimal_inflaton} we discuss specific models with
$V(\phi)=\frac{\lambda}{4}(\phi^2-v^2)^2$.
In Sec.~\ref{sec:annihilation} we study reheating via pair annihilation
of inflatons and compare that to the gravitational decay.
For concreteness, we shall consider single-field inflation models,
and assume $\Lambda\sim 1$ TeV and $N\sim 10^{32}$ throughout this paper
unless stated otherwise. 
We work with the metric signature $(-+++)$.

\section{enhanced decay of inflaton}\label{sec:grav_decay}
In theories with non-minimal couplings between the Ricci curvature and
scalar fields, e.g., supergravity, $\mathcal{R}^2$ gravity,
scalar-tensor gravity, and higher dimensional gravity theories, inflaton
fields can decay via gravitational effects. 

The perturbative decay rate, $\Gamma_{\rm grav}$, is
typically given by 
\begin{eqnarray}
\label{grav_decay}
\Gamma_{\rm grav} \sim NC\frac{m_{\sigma}^3}{M_{\rm Pl}^2},
\end{eqnarray}
where $C$ is a model-dependent fudge factor.
Although the gravitational decay rate is suppressed by the Planck scale,
the large number of species, $N$, would compensate it. 
One usually takes $N\sim 10^2-10^3$ and $NC\sim \mathcal{O}(1)$.

In Dvali's scenario, $N\sim 10^{32}$, and 
$M_{\rm Pl}^2$ is bounded from below by
Eq.~(\ref{dvali_bound}). The decay rate is therefore bounded from above as
\begin{eqnarray}
\Gamma_{\rm grav}  \lesssim C\frac{m_{\sigma}^3}{\Lambda^2}.
\end{eqnarray}
When Dvali's bound  is saturated, the decay
proceeds very fast and produces radiation\footnote{The ``radiation'' may
contain both visible and hidden sectors.} and entropy in the universe
efficiently.  

But, what if the decay is too efficient, and too much radiation is
produced? This is the argument that we shall use throughout this paper.

The most conservative, and model-independent constraint on Dvali's
proposal can be obtained by the following argument: at any epoch during
or after inflation, the energy density of the universe must not exceed
the Planck energy density. 

If the energy density of inflaton during inflation is less than the
Planck energy density, $\rho_{\rm inf}<M_{\rm Pl}^4$, energy conservation
demands that 
the energy density of radiation must also satisfy 
$\rho_{\rm rad}< \rho_{\rm inf} < M_{\rm Pl}^4$. 

The expansion rate of the universe during reheating roughly equals the
total decay rate of inflaton, $H(t_{\rm rh})\sim \Gamma_{\rm tot}\gtrsim
\Gamma_{\rm grav}$. From the Friedmann equation and
Eq.~(\ref{grav_decay}), we get $(\Gamma_{\rm grav}M_{\rm Pl})^2\sim
\left(NCm_{\sigma}^3/M_{\rm Pl}\right)^2 \lesssim \rho_{\rm rad} <
M_{\rm Pl}^4$. Solving this inequality for inflaton mass, we get an
upper limit on the inflaton mass,
\begin{eqnarray}
\label{rho_rad}
m_{\sigma}<10^{8}\;{\rm GeV}\left(\frac{10^{30}}{NC}\right)^{1/3}.
\end{eqnarray}

This constraint is not very interesting in the conventional scenario in
which $NC\sim \mathcal{O}(1)$, e.g., $m_\sigma<10^{18}$~GeV.

However, in a large-$N$ limit, say $NC\sim 10^{30}$, the constraint
becomes very tight: $m_{\sigma}<10^8~{\rm GeV}\sim 10^{-10}M_{\rm Pl}$,
which is significantly tighter than the usual fine-tuning of the
inflaton mass, $m_\sigma\sim 10^{-6}M_{\rm Pl}$, for a successful
chaotic inflation model with $V(\phi)=m^2_\sigma\phi^2/2$.  
Therefore, Dvali's proposal is consistent only when the inflaton is very
light, significantly lighter than the conventional case.
While $m_{\sigma}$ is fined-tuned with respect to $M_{\rm Pl}$ (or $M_{\rm grav}$), it would be natural to have $m_{\sigma}\sim M_{\rm UV} \sim O$(TeV) whose value is consistent with Eq.~(\ref{rho_rad}).

A more 
powerful constraint comes from the expansion
rate of the universe during inflation, $H_{\rm inf}$.
The energy density of inflaton during inflation is related to
$H_{\rm inf}$ via the Friedmann equation, $\rho_{\rm inf}=3M_{\rm
Pl}^2H^2_{\rm inf}$. 
Energy conservation then demands that 
the energy density of radiation must satisfy $\rho_{\rm rad}
<\rho_{\rm inf}=3M_{\rm Pl}^2H^2_{\rm inf}$, which yields
a bound on the gravitational decay rate, $\Gamma_{\rm
grav}<\sqrt{3}H_{\rm inf}$.

This bound may also be found as follows: 
since the expansion rate decreases as inflation ends, the expansion rate
during inflation, $H_{\rm inf}$, is greater than that during reheating:
$H_{\rm
inf}>H(t_{\rm rh})\sim\Gamma_{\rm tot}\gtrsim\Gamma_{\rm grav}$.
We can use this inequality to constrain $\Gamma_{\rm grav}$, if we know
what $H_{\rm inf}$ is.

How do we constrain $H_{\rm inf}$ observationally?
The amplitude of primordial
gravity waves is related to the expansion rate of the universe during
inflation as 
\begin{equation}
 H_{\rm inf}^2\simeq\frac{\pi^2 M_{\rm Pl}^2\Delta_h^2(k)}2
=\frac{\pi^2 M_{\rm Pl}^2r\Delta_{\mathcal R}^2(k)}2,
\end{equation}
where $\Delta_{h}^2(k)$ and $\Delta_{\mathcal R}^2(k)$ are the
dimensionless power spectrum of tensor and curvature perturbations,
respectively \cite{verde/etal:03}. 
The current observational constraint on the tensor-scalar ratio, $r\equiv
\Delta_h^2(k)/\Delta_{\mathcal R}^2(k)$, is 
$r\lesssim 1$, and the curvature perturbation is of order $\Delta_{\mathcal
R}^2(k)\sim 2\times 10^{-9}$ \cite{page/etal:07,spergel/etal:07}. 

Combining these with Eq.~(\ref{grav_decay}), we get a tighter constraint,
\begin{eqnarray}
m_{\sigma}
<10^{7}\;{\rm
 GeV}\left(\frac{10^{30}}{NC}\right)^{1/3}\left(\frac{r\Delta_{\mathcal
      R}^2}{2\times10^{-9}}\right)^{1/6}.\label{eq:H_inf} 
\end{eqnarray}
Note that this limit is also fairly model-independent,
and makes the fine-tuning of the inflaton mass even tighter.

As this constraint is insensitive to the precise value of $r$ or
$\Delta^2_{\mathcal R}$, the future observations of the B-mode
polarization of the cosmic microwave background, which would reach
$r\sim \mathcal{O}(10^{-2})$, will not improve the
constraint significantly.

One can obtain even tighter constraints by considering the following
limit on the reheating temperature,
\begin{eqnarray}
1\;{\rm MeV}\lesssim T_{\rm rh}\lesssim 10^8\;{\rm GeV},\nonumber
\end{eqnarray}
where the lower bound comes from the successful primordial
nucleosynthesis and the upper bound comes from the requirement that
thermal overproduction of gravitino/moduli be
avoided \cite{pagels/primack:82, coughlan/etal:83, ellis/etal:86}.
While the lower bound on the temperature must be satisfied for any
models\footnote{Here, we assume that $N$-species fields are unharmful
and cascade into radiation in the visible sector, or stable dark matter
particles, eventually. If 
$N$-species fields are long-lived, the 
reheat temperature is roughly given by $T_{\rm rh}\sim \sqrt{M_{\rm
Pl}}(\Gamma_{\sigma}^{-1}+\Gamma_{N}^{-1})^{-1/2}$, where
$\Gamma_{\sigma}$ is the total decay rate of inflaton and $\Gamma_{N}$
is that of $N$-species fields. Long-lived, but unstable, $N$-species
fields might 
cause problems, in a way similar to the late decay of moduli
\cite{coughlan/etal:83}.}, the upper bound is a model-dependent limit. 
Therefore, this limit is not as generic as the previous two
limits. Nevertheless, the resulting constraint on the inflaton mass is
the strongest, as we shall show below.

If the gravitational decay is a dominant process ($\Gamma_{\rm tot}\sim
\Gamma_{\rm grav}$), we get $NT_{\rm rh}^4\sim (\Gamma_{\rm tot}M_{\rm
Pl})^2\sim \left(NCm_{\sigma}^3/M_{\rm Pl}\right)^2$, where we have used
the Friedmann equation and Eq.(\ref{grav_decay}); thus, 
the reheat temperature is given by  
\begin{eqnarray}
T_{\rm rh}\sim N^{1/4}\sqrt{\frac{Cm_{\sigma}^3}{M_{\rm
 Pl}}},\label{eq:reheat} 
\end{eqnarray}
and the above limit on $T_{\rm rh}$ from the nucleosynthesis and
gravitino/moduli problem yields
\begin{eqnarray}
0.1\;{\rm GeV} <
 m_{\sigma}\left(\frac{NC^2}{10^{30}}\right)^{1/6} <
 10^{6.3}\;{\rm GeV},\label{reheat_temperature}
\end{eqnarray}
which can be tighter than Eqs.~(\ref{rho_rad}) and (\ref{eq:H_inf}), depending on $C$.

Even if the decay process is dominated by non-gravitational ones
(i.e. direct interactions), the upper bound is still valid, as the
gravitational decay channel gives the minimal decay rate. The lower bound is
a necessary condition to reheat  the universe after inflation mainly by
gravitational decay of inflaton. 

At the earlier stage of reheating, there may be non-perturbative decay of
inflaton via preheating, depending on the magnitude of direct interactions.
Our argument is valid even after preheating, if any, as the
gravitational decay channel gives the minimal decay rate in any case.

Here, we have used a simple, but rather crude, argument to make the main
point of this paper. There is still one unknown quantity, $C$, which
depends on specific models. How do we determine $C$? We shall present
more concrete models in the following sections.

\subsection{Decay induced by $f(\phi)R$ gravity}
In this section we use $C$ that we have derived in
\cite{watanabe/komatsu:07}.  

Almost all candidate theories of fundamental physics that involve
some compactification of the extra spatial dimensions are expected to
yield $f(\phi)R$ term, instead of the Einstein-Hilbert term,  in the
action, the form of $f(\phi)$ depending on models.  

The gravitational decay rate of inflaton into all the species that could
have existed at the reheating epoch is then given by  
$C=[F_1(v)]^2/(128\pi M_{\rm Pl}^2)$ \cite{watanabe/komatsu:07}, or
\begin{eqnarray}
\Gamma_{\rm tot}\simeq N\frac{[F_1(v)]^2}{128\pi M_{\rm Pl}^2}\frac{m_{\sigma}^3}{M_{\rm Pl}^2},\label{eq:decay}
\end{eqnarray}
where $F_1(v)\equiv |f'(v)|\left[1+\frac32(f'(v)/M_{\rm Pl})^2\right]^{-1/2}$, $\left.f'(v)\equiv \partial f/\partial\phi\right|_{\phi=v}$, and
$v\equiv \langle\phi\rangle$ is the vacuum expectation value of $\phi$. Here,
$N$-species fields are scalars
that are minimally coupled to gravity with a single mass scale,
$\Lambda$, and $\Lambda\ll m_\sigma$.\footnote{While we consider scalar matter
(bosons) only,  
one can calculate $C$ for fermions as well.
However, the gravitational decay channel to those light (compared to inflaton)
fermions is suppressed by their mass, as massless fermions are
conformally coupled to gravity\cite{watanabe/komatsu:07}.} 
Of course, all the fields do not need to have exactly the same mass, and
our argument still applies when they have a moderate mass spectrum.  

In this model the inequality Eq.(\ref{rho_rad}) becomes 
\begin{eqnarray}
m_{\sigma}<10^8\;{\rm
GeV}\left(\frac{10^{32}}{N}\right)^{1/3}\left(\frac{M_{\rm
     Pl}}{F_1(v)}\right)^{2/3}.\label{rho_rad_2}  
\end{eqnarray}

To make the constraint on $m_\sigma$ slightly more general, let us
parametrize $N$ in terms of $\alpha$ as
$\alpha\equiv N\Lambda^2/M_{\rm Pl}^{2}\lesssim 1$. 
In order to solve the hierarchy problem with Dvali's argument,
$\alpha\sim 1$ is required.
Note that the minimum of this parameter is given by
$\alpha_{\rm grav}=\Lambda^2/M_{\rm Pl}^2$, which represents weakness of
gravity for particles with mass of $\Lambda$.  

Assuming a typical value of $|f'(v)|\sim M_{\rm Pl}$, we find that 
the inflaton mass must be tuned to be smaller than $10^8\alpha^{-1/3}\; {\rm
GeV}$.  
This constraint is most stringent when the hierarchy problem is solved
(i.e. $\alpha\sim 1$). 

One may reverse the argument by taking the inflaton
mass to be a typical value of chaotic inflation, $m_{\sigma}\sim
10^{12}$~GeV, which limits $f'(v)$ as $F_1(v)/M_{\rm
Pl}\sim |f'(v)|/M_{\rm Pl}<10^{-6}\alpha^{-1/2}$, i.e., $f'(v)$ must be fine-tuned.

The expansion rate during inflation gives  a stronger constraint
[Eq.~(\ref{eq:H_inf})]:
\begin{eqnarray}
m_{\sigma}<10^7\;{\rm GeV}\left(\frac{10^{32}}{N}\right)^{1/3}\left(\frac{M_{\rm Pl}}{F_1(v)}\right)^{2/3}\left(\frac{r\Delta_{\mathcal R}^2}{2\times10^{-9}}\right)^{1/6}.\nonumber
\end{eqnarray}
For $|f'(v)|\sim M_{\rm Pl}$, $m_\sigma<10^7\alpha^{-1/3}\; {\rm GeV}$.
For $m_\sigma\sim 10^{12}$~GeV, $F_1(v)/M_{\rm Pl}\sim |f'(v)|/M_{\rm Pl}<10^{-7.5}\alpha^{-1/2}$.

The limit on the reheating temperature [Eq.~(\ref{eq:reheat})]
yields even stronger constraint [Eq.~(\ref{reheat_temperature})]:
\begin{eqnarray}
0.3\;{\rm GeV}< m_{\sigma}\left(\frac{N}{10^{32}}\right)^{1/6}\left(\frac{F_1(v)}{M_{\rm Pl}}\right)^{2/3} < 10^{6.8}\;{\rm GeV}.\nonumber
\end{eqnarray}
For $|f'(v)|\sim M_{\rm Pl}$, the upper limit on the mass is
$m_\sigma<10^{6.8}\alpha^{-1/6}$~GeV. 
For $m_{\sigma}\sim 10^{12}$~GeV,
$F_1(v)/M_{\rm Pl}\sim |f'(v)|/M_{\rm Pl}< 10^{-8}\alpha^{-1/4}$.\footnote{The upper limit is
identical to Eq.~(18) in \cite{watanabe/komatsu:07}, when $\alpha\sim 10^{-32}$.} 

In summary, we have confirmed that, using a physically motivated form of
$C$, Dvali's large-$N$ species solution to the hierarchy problem demands
tight fine-tuning of $m_\sigma$ or $f'(v)$. But, how bad are these
fine-tunings? 

\section{Worked example: Ginzburg-Landau
 potential}\label{sec:nonminimal_inflaton}

The precise values of $m_{\sigma}$ and
$f'(v)$ depend on models. We shall study this point in further detail,
by using a specific model given by the following action:
\begin{eqnarray}
\mathcal{L}&=&\sqrt{-g}\left[\frac{1}{2}f(\phi)R-\frac{1}{2} \partial_{\mu}\phi\partial^{\mu}\phi-V(\phi) \right]+\mathcal{L}_{\rm matter},\nonumber\\
& &f(\phi)= M_{\rm Pl}^2 +\xi(\phi^2-v^2),\\
& &V(\phi)=\frac{\lambda}{4}(\phi^2-v^2)^2,\nonumber
\end{eqnarray}
where the form of $f(\phi)$ is a popular non-minimal
 coupling\footnote{In our notation, the conformal coupling corresponds
 to $\xi=-1/6$.} with 
 the condition, $f(v)=M_{\rm Pl}^2$, that recovers General
Relativity after inflation \cite{watanabe/komatsu:07,
SalopekBondBardeen89, kaiser:95}. 

The form of $V(\phi)$ is also a popular Ginzburg-Landau-type potential.
Therefore, our example is not exotic or peculiar; on the contrary, this
 is one of the best studied case, and therefore we hope that this
 example helps ones understand better how much fine-tuning is required
 by Dvali's solution to the hierarchy problem.

We now consider what happens when inflaton leaves the slow-roll regime,
and begins to oscillate around its potential minimum. The inflaton
quanta from the oscillations decay into relativistic $N$ species at least
perturbatively and gravitationally. 
The inflaton mass is given by the curvature of potential around the minimum,
$m_{\sigma}^2\equiv\partial^2V(\phi)/\partial\phi^2|_{\phi=v}=2\lambda
v^2 $.

We use the most conservative (probably overly conservative) limit on the
 radiation energy, $\rho_{\rm rad}<M_{\rm Pl}^4$
 [Eq.~(\ref{rho_rad_2})], to find a limit on  $\lambda$ and $|\xi|$ as
 \begin{eqnarray}
\lambda  < \frac{10^{-20}}{|\xi|^{4/3}}\left(\frac{10^{32}}{N}\right)^{2/3}\left(\frac{M_{\rm Pl}}{v}\right)^{10/3}\left(1+6\xi^2\frac{v^2}{M_{\rm Pl}^2}\right)^{2/3},\label{eq:lambda}
\end{eqnarray}
which is extremely tight, compared to the existing constraints from
inflation \cite{futamase/maeda:89, makino/sasaki:91, fakir/unruh:90:92,
kaiser:95, hwang/noh:98, komatsu/futamase:99,
tsujikawa/gumjudpai:04}. In fact, this limit excludes most of the
parameter space allowed by the WMAP 3-yr data.
One can find even stronger constraints by considering the expansion rate
during inflation, or the limits on the reheating temperature from the 
gravitino/moduli problem.

It follows from Eq.~(\ref{eq:lambda}) that
it is difficult to avoid fine-tuning of one parameter without
fine-tuning the other parameters. Either $\lambda$, $\xi$, or $v/M_{\rm
Pl}$, or perhaps all of them, need to be fine-tuned for the 
large-$N$ species solution to the hierarchy problem to
be consistent with inflationary cosmology.

\section{Reheating by Pair annihilation of inflatons}
\label{sec:annihilation}
So far, we have studied implications of the gravitational decay of
inflatons enhanced by the existence of large-$N$ 
species. In deriving Eq.~(\ref{grav_decay}), we assume that inflatons 
decay into $N$-species fields via an effective trilinear vertex,
$\mathcal{L}_{\rm int}\propto \sigma\chi\chi$, which
couples inflaton quanta, $\sigma$, to a pair of $N$ species,
$\chi$. However, what if such a coupling 
is forbidden by the symmetry of the inflaton field? Are there any other
channels to reheat the universe after inflation?  

In theories of $f(\phi)R$ gravity, the effective interaction Lagrangian
is given by the Taylor series expansion of $f(\phi)$ around the vacuum
expectation value of $\phi$, $\phi=v+\sigma$, where $\sigma$ is the
inflaton quanta. We find \cite{watanabe/komatsu:07}
\begin{eqnarray}
\nonumber
 \frac{\mathcal{L}_{\rm int}}{\sqrt{-g}} &=& \frac{f'(v)}{M_{\rm Pl}^2} 
\left[\sigma U(\chi) - (\partial_\mu \sigma) G^\mu(\chi)\right]\\
& &
\nonumber
+  \frac{f''(v)}{2M_{\rm Pl}^2}
\left[\sigma^2 U(\chi) - 2\sigma (\partial_\mu \sigma)G^\mu(\chi)\right]\\ 
& &
\nonumber
+ \frac{[f'(v)]^2}{2M_{\rm Pl}^4}
\left[ 2\sigma (\partial_\mu \sigma)G^\mu(\chi)
-(\partial_\mu \sigma) (\partial_\nu \sigma) H^{\mu\nu}(\chi)\right]\\
& & +\frac{f'''(v)}{6M_{\rm Pl}^2}\left[\sigma^3 U(\chi)+\dots\right]+\dots,
\end{eqnarray}
where $U(\chi)$ is the scalar field potential, e.g.,
$U(\chi)=m_\chi^2\chi^2/2+\lambda\chi^4/4+\dots$, and the other
functions are given by
\begin{eqnarray}
 G^\mu&=& \frac12g^{\mu\alpha}\chi(\partial_\alpha\chi), \\
 H^{\mu\nu}&=& \frac14g^{\mu\nu}\chi^2.
\end{eqnarray}
Note that one can also derive the interaction Lagrangian for fermions
systematically in a similar manner. 

The first terms in ${\mathcal L}_{\rm int}$ that are proportional to
$f'(v)$ yield the decay of 
$\sigma$ with the rate given by Eq.~(\ref{eq:decay}), whereas those
proportional to $f''(v)$ and $[f'(v)]^2$ yield the pair annihilation.

Now, let us imagine that the first derivative of $f(\phi)$ vanishes at
the vacuum expectation value, $f'(v)=0$, which will shut off the decay
channel. We also assume that $\chi$
are massive free fields, $U(\chi)=m_\chi^2\chi^2/2$. 
The interaction Lagrangian at
the lowest order in $\sigma$ for this case is given by
\begin{equation}
\nonumber
 \frac{\mathcal{L}_{\rm int}}{\sqrt{-g}} =
  \frac{f''(v)}{2M_{\rm Pl}^2}
\left[\frac12m_\chi^2\sigma^2\chi^2 - g^{\mu\nu}
\sigma (\partial_\mu \sigma)\chi(\partial_\nu\chi)\right],
\end{equation}
which may also be written in the form of 
$\hat{g}^2\hat{\sigma}^2\chi^2$ (after some integration by parts and the use of
equation of motion, $\Box\chi=\partial U/\partial\chi$), where
$\hat{g}^2=f''(v)\left[1+\frac32(f'(v)/M_{\rm Pl})^2\right]^{-1}(m_{\chi}^2+s/2)/(4M_{\rm Pl}^2)$
and $s\equiv - g_{\mu\nu}(q_1^\mu+q_2^\mu)(q_1^\nu+q_2^\nu)$ is the
square of the total initial 
4-momentum of incoming inflaton quanta.
We must cannonically normalize inflaton quanta as $\hat{\sigma}\equiv \sigma\sqrt{1+\frac32(f'(v)/M_{\rm Pl})^2}$.
We calculate the annihilation cross section from this interaction Lagrangian as 
\begin{eqnarray}
\sigma_{\rm ann}
=N\frac{[F_2(v)]^2}{32\pi M_{\rm Pl}^4}\frac1{s}\left(m_{\chi}^2+\frac{s}2\right)^2\sqrt{\frac{s-4m_{\chi}^2}{s-4m_{\sigma}^2}},
\end{eqnarray}
where $F_2(v)\equiv |f''(v)|\left[1+\frac32(f'(v)/M_{\rm Pl})^2\right]^{-1}$.

The annihilation rate of inflatons is then given by
\begin{eqnarray}
\nonumber
\Gamma_{\rm ann}&\equiv& n_{\sigma}\langle\sigma_{\rm ann}v_{\rm rel}\rangle \\
&\simeq& N\frac{3[F_2(v)]^2H_{\rm rh}^2\langle s\rangle}{64\pi M_{\rm
 Pl}^2m_{\sigma}}
\geq N\frac{3[F_2(v)]^2H_{\rm rh}^2m_\sigma}{16\pi M_{\rm
 Pl}^2},\label{eq:annihilation} 
\end{eqnarray}
where $n_{\sigma}=\rho_{\sigma}/m_{\sigma}\simeq 3M_{\rm Pl}^2H_{\rm
rh}^2/m_{\sigma}$ is the number density of inflaton quanta. 
In deriving Eq.~(\ref{eq:annihilation}) we have assumed 
$m_{\chi}\ll m_{\sigma}$. The relative velocity, $v_{\rm rel}$, is given by
$v_{\rm rel}=2\sqrt{1-4m_\sigma^2/s}$
in the center of mass frame.
Finally, the average of $s$, $\langle s\rangle$, is bounded from below,
$\langle s\rangle\geq 4m_\sigma^2$, where the equality is satisfied when
the inflaton quanta are at rest. While we expect them to be non-relativistic
at the beginning of reheating, $\langle s\rangle\sim 4m_\sigma^2$, we
keep the inequality explicitly in the following discussion.

In order for annihilation to be efficient during reheating, the
annihilation rate has to be greater than the expansion rate during
reheating, $\Gamma_{\rm ann}>H_{\rm rh}$. This is satisfied when
\begin{eqnarray}
\frac{\langle s\rangle}{4m_{\sigma}}
> \frac{10^{-7}{\rm GeV}}{[F_2(v)]^2}\left(\frac{10^{-6}M_{\rm
				      Pl}}{H_{\rm
				      rh}}\right)\left(\frac{10^{32}}{N}\right),
\end{eqnarray}
which is a rather weak lower bound on $\langle s\rangle/4m_{\sigma}$, 
which is approximately equal to $m_\sigma$ in the
non-relativistic limit, for $N\sim
10^{32}$. Therefore, the presence of large-$N$ species makes annihilation
very efficient. 

Let us compare $\Gamma_{\rm ann}$ [Eq.~(\ref{eq:annihilation})] with the
decay rate, $\Gamma_{\rm 
decay}$ [Eq.~(\ref{eq:decay})]:
\begin{eqnarray}
\frac{\Gamma_{\rm ann}}{\Gamma_{\rm decay}}\gtrsim
 24\left(\frac{F_2(v)M_{\rm Pl}}{F_1(v)}\right)^2\left(\frac{H_{\rm
    rh}}{m_{\sigma}}\right)^2,
\end{eqnarray}
where the approximate equality is satisfied when inflaton quanta are
non-relativistic.
We therefore find that the annihilation channel is not necessarily
smaller than the decay channel. It may be more informative to write this
result in the following form:
\begin{eqnarray}
\frac{\Gamma_{\rm ann}}{\Gamma_{\rm decay}}\gtrsim
 8\left(\frac{F_2(v)M_{\rm Pl}}{F_1(v)}\right)^2
\frac{n_\sigma}{M_{\rm Pl}^2m_\sigma}.
\end{eqnarray}
Thus, there is a critical number density above which the annihilation
channel dominates over the decay channel:
\begin{equation}
 n_\sigma^{\rm crit} \equiv \frac{M_{\rm Pl}^2m_\sigma}8 
\left(\frac{F_1(v)}{F_2(v)M_{\rm Pl}}\right)^2.
\end{equation}
For $f(\phi)=M_{\rm Pl}^2+\xi(\phi^2-v^2)$, for instance, the critical
density is given by
$n_\sigma^{\rm crit} = M_{\rm Pl}^2m_\sigma
({v}/{M_{\rm Pl}})^2\left[1+6\xi^2(v/M_{\rm Pl})^2\right]/8$.

\section{conclusions}
We have studied consistency between Dvali's large-$N$ species solution to
the gauge hierarchy problem and inflationary cosmology. 

If there exist
the large-$N$ species, the inflaton quanta decay or annihilate 
too efficiently, and reheat
the universe too much. We have found that, in order for this scenario to
produce successful reheating of the universe, either inflaton mass, vacuum
expectation value, or non-minimal gravitational coupling, or all of them,
must be fine-tuned to suppress the gravitational decay and annihilation
of the inflaton quanta.  

We have shown that fine-tuning of an extreme magnitude,
Eq.~(\ref{eq:lambda}), is required by using a widely-studied example. 
The constraint we have found indeed excludes most of parameter space of
the model. This example demonstrates that one must always check
whether reheating is successful, whenever their models contain
non-minimal coupling, such as $f(\phi)R$ gravity.

One may repeat the same analysis for supergravity inflation models
\cite{endo/etal:06, endo/etal:07, endo/kadota/etal:07} that contain not
only non-minimal gravitational coupling (the K\"ahler potential determines
the function $f(\phi)$), but also direct coupling terms in the
supergravity frame. However, as supersymmetry alone is able to solve 
 the gauge hierarchy problem (albeit $\mu$ problem still remains), 
 it seems difficult to motivate our having
 both supersymmetry and large-$N$ species (see, however \cite{berera/kephart:99}). 

Finally, let us point out the limitation and caveat of our analysis.

The constraints given in this paper are based exclusively upon
non-minimal gravitational couplings of inflaton. Therefore, if the
non-minimal coupling is 
totally absent, $f(\phi)\equiv M_{\rm Pl}^2$, or matter fields are
conformally coupled to gravity (e.g., massless scalars with $\xi=-1/6$,
massless fermions, etc), both classically and quantum
mechanically, reheating of the universe must be achieved by direct 
couplings between inflaton and matter fields. Our limits
on Dvali's scenario do not apply to such cases.

We have assumed that the cut-off scale for gravity is much higher
than that for the particle physics, $\Lambda$, and thus 
ignored quantum gravitational effects on the decay rates. 
It would be interesting to extend our analysis to the case where
the cut-off for gravity is also similar to $\Lambda$ \cite{ADD} or even lower than $\Lambda$ \cite{dvali/etal:01}.  

\begin{acknowledgments}
We would like to thank Arjun Berera, Jim Cline, Gia Dvali, Nemanja Kaloper, and Christos Kokorelis for comments on the paper.
Y.W. thanks Donghui Jeong and Jun Koda for discussions. 
E.K. acknowledges support from the Alfred P. Sloan Foundation.
\end{acknowledgments}

\end{document}